\documentclass[twocolumn,prl,superscriptaddress,showpacs,preprintnumbers,amsmath,amssymb]{revtex4}
\usepackage[ansinew,latin1]{inputenc}
\usepackage{psfrag}
\usepackage{array}
\usepackage{amssymb}
\usepackage{amsmath}
\usepackage{graphicx}

\def\kop{\tilde{\kappa}_{o+}}

\def\lsim{\mathrel{\rlap{\lower3pt\hbox{$\sim$}}
   \raise2pt\hbox{$<$}}}
\def\gsim{\mathrel{\rlap{\lower3pt\hbox{$\sim$}}
   \raise2pt\hbox{$>$}}}
\def\sqr#1#2{{\vcenter{\vbox{\hrule height.#2pt
        \hbox{\vrule width.#2pt height#1pt \kern#1pt
        \vrule width.#2pt}
        \hrule height.#2pt}}}}

\def\etal{{\it et al.}}

\newcommand{\beq}[1]{\begin{equation}\label{#1}}
\newcommand{\eeq}{\end{equation}}
\newcommand{\bea}[1]{\begin{eqnarray}\label{#1}}
\newcommand{\eea}{\end{eqnarray}}
\newcommand{\ba}{\begin{array}}
\newcommand{\ea}{\end{array}}

\newcommand{\rf}[1]{(\ref{#1})}

\begin{document}

\title{Limits on light-speed anisotropies 
from Compton scattering of high-energy electrons}

\author{J.-P.~Bocquet}
\affiliation{LPSC, UJF Grenoble 1, CNRS/IN2P3, INPG, 53 avenue des Martyrs 38026 Grenoble, France}
\author{D.~Moricciani}
\affiliation{INFN Sezione di Roma TV, 00133 Roma, Italy}
\author{V.~Bellini}
\affiliation{INFN Sezione di Catania and Universit\`a di Catania, 95100 Catania, Italy}
\author{M.~Beretta}
\affiliation{INFN Laboratori Nazionali di Frascati, 00044 Frascati,~Italy}
\author{L.~Casano}
\affiliation{INFN Sezione di Roma TV, 00133 Roma, Italy}
\author{A.~D'Angelo}
\affiliation{INFN Sezione di Roma TV and Universit\`a di Roma ``Tor Vergata," 00133 Roma, Italy}
\author{R.~Di~Salvo}
\affiliation{INFN Sezione di Roma TV, 00133 Roma, Italy}
\author{A.~Fantini}
\affiliation{INFN Sezione di Roma TV and Universit\`a di Roma ``Tor Vergata," 00133 Roma, Italy}
\author{D.~Franco}
\affiliation{INFN Sezione di Roma TV and Universit\`a di Roma ``Tor Vergata," 00133 Roma, Italy}
\author{G.~Gervino}
\affiliation{INFN Sezione di Torino  and Universit\`a di Torino, 10125 Torino, Italy}
\author{F.~Ghio}
\affiliation{INFN Sezione di Roma I and Istituto Superiore di Sanit\`a, 00161 Roma, Italy}
\author{G.~Giardina}
\affiliation{INFN Sezione di Catania and Universit\`a di Messina, 98166 Messina, Italy}
\author{B.~Girolami}
\affiliation{INFN Sezione di Roma I and Istituto Superiore di Sanit\`a, 00161 Roma, Italy}
\author{A.~Giusa}
\affiliation{INFN Sezione di Catania and Universit\`a di Catania, 95100 Catania, Italy}
\author{V.G.~Gurzadyan}
\affiliation{Yerevan Physics Institute, 375036 Yerevan, Armenia}
\affiliation{Yerevan State University, 375025 Yerevan, Armenia}
\author{A.~Kashin}
\affiliation{Yerevan Physics Institute, 375036 Yerevan, Armenia}
\author{S.~Knyazyan}
\affiliation{Yerevan Physics Institute, 375036 Yerevan, Armenia}
\author{A.~Lapik}
\affiliation{Institute for Nuclear Research, 117312 Moscow, Russia}
\author{R.~Lehnert}
\email[]{ralf.lehnert@nucleares.unam.mx}
\affiliation{ICN,
Universidad Nacional Aut\'onoma de M\'exico,
A.~Postal 70-543, 04510 M\'exico D.F., Mexico}
\author{P.~Levi~Sandri}
\affiliation{INFN Laboratori Nazionali di Frascati, 00044 Frascati,~Italy}
\author{A.~Lleres}
\affiliation{LPSC, UJF Grenoble 1, CNRS/IN2P3, INPG, 53 avenue des Martyrs 38026 Grenoble, France}
\author{F.~Mammoliti}
\affiliation{INFN Sezione di Catania and Universit\`a di Catania, 95100 Catania, Italy}
\author{G.~Mandaglio}
\affiliation{INFN Sezione di Catania and Universit\`a di Messina, 98166 Messina, Italy}
\author{M.~Manganaro}
\affiliation{INFN Sezione di Catania and Universit\`a di Messina, 98166 Messina, Italy}
\author{A.~Margarian}
\affiliation{Yerevan Physics Institute, 375036 Yerevan, Armenia}
\author{S.~Mehrabyan}
\affiliation{Yerevan Physics Institute, 375036 Yerevan, Armenia}
\author{R.~Messi}
\affiliation{INFN Sezione di Roma TV and Universit\`a di Roma ``Tor Vergata," 00133 Roma, Italy}
\author{V.~Nedorezov}
\affiliation{Institute for Nuclear Research, 117312 Moscow, Russia}
\author{C.~Perrin}
\affiliation{LPSC, UJF Grenoble 1, CNRS/IN2P3, INPG, 53 avenue des Martyrs 38026 Grenoble, France}
\author{C.~Randieri}
\affiliation{INFN Sezione di Catania and Universit\`a di Catania, 95100 Catania, Italy}
\author{D.~Rebreyend}
\email[]{rebreyend@lpsc.in2p3.fr}
\affiliation{LPSC, UJF Grenoble 1, CNRS/IN2P3, INPG, 53 avenue des Martyrs 38026 Grenoble, France}
\author{N.~Rudnev}
\affiliation{Institute for Nuclear Research, 117312 Moscow, Russia}
\author{G.~Russo}
\affiliation{INFN Sezione di Catania and Universit\`a di Catania, 95100 Catania, Italy}
\author{C.~Schaerf}
\affiliation{INFN Sezione di Roma TV and Universit\`a di Roma ``Tor Vergata," 00133 Roma, Italy}
\author{M.L.~Sperduto}
\affiliation{INFN Sezione di Catania and Universit\`a di Catania, 95100 Catania, Italy}
\author{M.C.~Sutera}
\affiliation{INFN Sezione di Catania and Universit\`a di Catania, 95100 Catania, Italy}
\author{A.~Turinge}
\affiliation{Institute for Nuclear Research, 117312 Moscow, Russia}
\author{V.~Vegna}
\affiliation{INFN Sezione di Roma TV and Universit\`a di Roma ``Tor Vergata," 00133 Roma, Italy}

\date{\today}

\begin{abstract}

The possibility of anisotropies in the speed of light 
relative to the limiting speed of electrons 
is considered. 
The absence of sidereal variations 
in the energy of Compton-edge photons
at the ESRF's GRAAL facility
constrains such anisotropies
representing the first non-threshold collision-kinematics study of Lorentz violation. 
When interpreted within the minimal Standard-Model Extension,
this result yields the two-sided limit of $1.6 \times 10^{-14}$  at $95\,\%$ confidence level
on a combination of the parity-violating
photon and electron coefficients 
$(\kop)^{YZ}$, 
$(\kop)^{ZX}$,
$c_{TX}$, and
$c_{TY}$.
This new constraint provides an improvement over previous bounds 
by one order of magnitude. 

\end{abstract}


\maketitle

Properties of the speed of light $c$, 
such as isotropy and constancy irrespective of the motion of the source, 
play a key role in physics. 
For example,
they are instrumental for both 
the conceptual foundations as well as the experimental verification 
of Special Relativity, 
and they currently provide the basis for the definition of length 
in the International System of Units. 
It follows that improved tests of, 
e.g., 
the isotropy of light propagation
remain of fundamental importance 
in physics.

Experimental searches for anisotropies in $c$
are further motivated by
theoretical studies 
in the context of quantum gravity:
it has recently been realized 
that a number of approaches to Planck-scale physics, 
such as strings, 
spacetime-foam models, 
non-commutative field theory, 
and varying scalars, 
can accommodate minuscule violations of Lorentz symmetry~\cite{lotsoftheory}. 
At presently attainable energies, 
such Lorentz-breaking effects 
can be described by the Standard-Model Extension (SME), 
an effective field theory 
that incorporates both 
the usual Standard Model and General Relativity 
as limiting cases~\cite{sme}. 
To date, 
the minimal Standard-Model Extension (mSME),
which contains only relevant and marginal operators, 
has provided the basis for numerous tests of Special Relativity 
in a wide variety of physical systems~\cite{cpt07,kr}.

In this work, 
we will study 
photons and electrons 
in an environment 
where gravity is negligible.
Lorentz violation 
is then described by
the single-flavor QED limit of the flat-spacetime mSME~\cite{sme,collider}.
This limit contains the real, spacetime-constant mSME coefficients 
$(k_{F})^{\mu\nu\rho\lambda}$, $(k_{AF})^{\mu}$, 
$b^\mu$, $c^{\mu\nu}$, $d^{\mu\nu}$, and $H^{\mu\nu}$,
which control the extent of different types of Lorentz and CPT violation.
Note, 
however, 
that $c^{\mu\nu}$ and $\tilde{k}^{\mu\nu}\equiv(k_F)_\alpha{}^{\mu\alpha\nu}$ 
are observationally indistinguishable 
in a photon--electron system: 
suitable coordinate rescalings 
freely transform the $\tilde{k}^{\mu\nu}$ and $c^{\mu\nu}$ parameters
into one another~\cite{km0102,collider}. 
Physically, 
this represents the fact 
that the speed of light is measured 
{\em relative} to the speed of electrons.
We exploit this freedom 
by selecting the specific scaling $c^{\mu\nu}=0$
in intermediate calculations.
However, 
we reinstate this coefficient 
in the final result 
for generality.

From a phenomenological perspective,
the dominant mSME coefficient is $k_F$, 
which causes a direction- and polarization-dependent speed of light~\cite{km0102}.
Various of its components 
have been tightly bounded with 
astrophysical observations~\cite{km06}, 
Michelson--Morley tests~\cite{kr,newMM},
and collider physics~\cite{collider}.
We will bound the $\kop$ piece of $\tilde{k}^{\mu\nu}$, 
which is an antisymmetric $3\times3$ matrix; 
it currently obeys the weakest limits, 
so all other mSME coefficients can be set to zero 
in what follows.
An mSME analysis then reveals that 
the photon's dispersion relation is modified:
\beq{ModDR}
\omega=(1-\vec{\kappa}\cdot\hat{\lambda})\,\lambda+{\mathcal O}(\kappa^2)\;.
\eeq
\noindent Here, $\lambda^\mu=(\omega,\lambda\hat{\lambda})$ denotes the photon 4-momentum
and $\hat{\lambda}$ 
is a unit 3-vector. The three components of $\kop$ 
have been assembled into the form of a 3-vector:
$\vec{\kappa}\equiv((\kop)^{23}, (\kop)^{31}, (\kop)^{12})
\equiv(\kappa_X,\kappa_Y,\kappa_Z)$.
This vector specifies a preferred direction, 
which violates Lorentz symmetry;
it can be interpreted 
as generating a direction-dependent refractive index of the vacuum 
$n(\hat{\lambda})\simeq1+\vec{\kappa}\cdot\hat{\lambda}$.
The electron's dispersion relation 
$E(p)=\sqrt{m^2+p^2}$
remains unaltered
with our choice of coordinate scaling.

The dispersion relation~\rf{ModDR} shows that
for a given $\lambda$, 
the photon energy $\omega$ 
depends on the direction $\hat{\lambda}$ of the photon's 3-momentum,
which exposes the $\kop$ anisotropies. 
Reversing the direction of 
$\hat{\lambda}$ establishes parity violation in these anisotropies.
The basic experimental idea is that
in a terrestrial laboratory 
the photon 3-momentum in a Compton-scattering process 
changes direction due to the Earth's rotation.  
The photons are thus affected by the anisotropies in Eq.~\rf{ModDR}
leading to sidereal effects in the kinematics of the process. 
Precision measurements of Compton scattering of ultrarelativistic electrons 
circulating in a high-energy accelerator 
could help reveal such effects.
In a previous paper~\cite{OGR}, 
we have established 
the high sensitivity of a method 
based on the analysis of the Compton-edge (CE) energy
(i.e., the minimal energy) 
of the scattered electrons 
using the GRAAL beamline~\cite{GRAAL} 
at the European Synchrotron Radiation Facility (ESRF, Grenoble, France).
The present work utilizes dedicated GRAAL data 
and an improved set-up to extract a competitive bound on $\kop$
at the $10^{-14}$ level.
Note that for the first time, 
kinematical non-threshold physics in a particle collision
is exploited to test Lorentz symmetry:
previous studies 
have employed predictions involving thresholds, 
such as shifts in thresholds 
(e.g., $\gamma\gamma\to e^+e^-$),
the presence of novel processes above certain thresholds
(e.g., photon decay), 
or the absence of conventional collisions above certain thresholds
(e.g., neutron stability)~\cite{thres}. 

The experimental set-up at GRAAL
involves counter-propagating incoming electrons and photons 
with 3-momenta $\vec{p}=p\,\hat{p}$ and $\vec{\lambda}=-\lambda\,\hat{p}$, 
respectively. The conventional CE then occurs
for outgoing photons that are backscattered at $180^\circ$,
so that the kinematics is essentially one dimensional
along the beam direction $\hat{p}$.
Energy conservation for this process reads
\beq{Econs}
E(p)+(1+\vec{\kappa}\cdot\hat{p})\,\lambda
=E(p-\lambda-\lambda')+(1-\vec{\kappa}\cdot\hat{p})\,\lambda'\;,
\eeq
where $\vec{\lambda}'=\lambda'\,\hat{p}$ is the 3-momentum
of the CE photon, 
and 3-momentum conservation has been implemented.
At leading order, 
the physical solution of Eq.~\rf{Econs} is
\beq{modCE}
\lambda'\simeq\lambda_{\rm CE}
\left[\,
1 + \frac{2\,\gamma^2}{(1 + 4\, \gamma\, \lambda\, /\, m)^2}\,\vec{\kappa}\cdot\hat{p}
\,\right]\,.
\eeq
Here, 
$\lambda_{\rm CE} = \frac{4\, \gamma^2 \, \lambda }{1 + 4\, \gamma\, \lambda\, /\, m}$ 
denotes the conventional value of the CE energy. 
Given the actual experimental data of 
$m=511\,$keV, 
$p=6030\,$MeV, 
and $\lambda=3.5\,$eV, 
yields $\gamma \simeq p/m = 11800$ and $\lambda_{\rm CE}=1473\,$MeV.
The numerical value of the factor in front of $\vec{\kappa}\cdot\hat{p}$ 
is about $1.6\times10^8$.
It is this large  amplification factor (essentially given 
by $\gamma^2$) that 
yields the exceptional sensitivity of the CE 
to $\kop$.

For comparisons between tests, 
mSME coefficients are conventionally specified 
in the Sun-centered inertial frame $(X,Y,Z)$~\cite{kr}, 
in which $\vec{\kappa}$ is constant, 
$\hat{Z}$ parallel to the Earth's rotation axis, 
and terrestrial laboratories rotate 
with a frequency $\Omega\simeq2\pi/(23\,$h$\;56\,$min$)$ 
about $\hat{Z}$. 
Hence, 
$\hat{p}(t)\simeq(0.9\cos \Omega t,0.9 \sin \Omega t,0.4)$,
where the beam direction of GRAAL has been used, 
and the time $t=0$ has been chosen appropriately.  
Then, Eq.~\rf{modCE} becomes
\beq{sidereal}
\lambda'\simeq\tilde{\lambda}_{\rm CE}+
0.91\,\frac{2\,\gamma^2\,\lambda_{\rm CE}}{(1 + 4\,\gamma\,\lambda\, / \,m )^2}
\sqrt{\kappa_X^2+\kappa_Y ^2}
\,\sin\Omega t\,.
\eeq
Here, 
we have absorbed the time-independent $\kappa_Z$ piece 
into 
$\tilde{\lambda}_{\rm CE}$ 
and dropped an irrelevant phase.
Equation~\rf{sidereal} clearly exposes the possibility of sidereal variations in the CE 
and represents our main theoretical result.

Small deviations $\delta\varphi$ from 
exactly counter-propagating incoming photons and electrons
as well as the Lorentz-violating dispersion relation~\rf{ModDR} itself
can place the CE direction
slightly off-axis. 
Moreover, 
Lorentz violation can affect laser lines as well as the spectrometer's
$B$ field.
These effects have been carefully studied 
and shown to be negligible, 
so the primary Lorentz-violating effect
in the present context is given by Eq.~\rf{sidereal}.

\begin{figure}[t]
\begin{center}
\includegraphics[width=0.9\hsize]{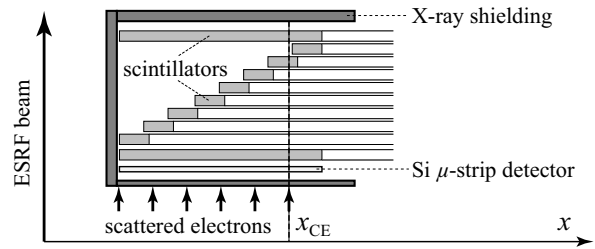}
\end{center}
\vskip-10pt
\caption{Schematic drawing of the tagging system.
\vspace{-5mm}}
\label{test}
\end{figure}

Incoming photons are generated by a high-power Ar laser
located about $40\,$m from the intersection region. 
The laser beam enters the vacuum via a MgF window 
and is then reflected by an Al-coated Be mirror 
towards the $6.03\,$GeV electron beam. 
The laser and electron beams overlap 
over a $6.5\,$m long straight section. 
Photons are finally absorbed in a four-quadrant calorimeter, 
which allows the stabilization of the laser-beam barycenter to $0.1\,$mm. 
This level of stability is necessary 
and corresponds to a major improvement of the set-up 
relative to our previous result. 
Due to their energy loss, 
scattered electrons are extracted from the main beam 
in the magnetic dipole 
following the straight section. 
Their position can then be accurately measured 
in the so-called tagging system (Fig.~\ref{test}) 
located $50\,$cm after the exit of the dipole. 
This system plays the role of a magnetic spectrometer
from which we can infer the electron momentum.
The tagging system is composed of a position-sensitive Si $\mu$-strip detector 
(128~strips of $300\,$$\mu$m pitch, $500\,$$\mu$m thick) 
associated to a set of fast plastic scintillators 
for timing information and triggering of the DAQ. 
These detectors are placed inside a movable box 
shielded against the huge X-ray background 
generated in the dipole. A spring system 
ensures that 
the position-sensitive Si detector 
always touches the box end.
As will be discussed later, 
the X-ray induced heat load 
is the origin of sizable variations in the box temperature, 
correlated with the ESRF beam intensity.
This produces 
a continuous drift of the detector 
due to the dilation of the box. 

\begin{figure}
\begin{center}
\includegraphics[width=0.9\hsize]{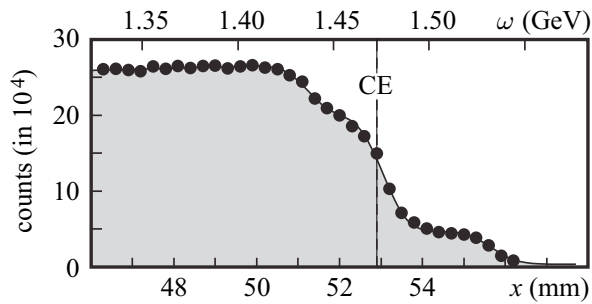}
\end{center}
\vskip-10pt
\caption{Si $\mu$-strip count spectrum near the CE 
and the fitting function (see text) vs.\ position $x$
and photon energy $\omega$. 
The three edges corresponding to the lines
$364$, $351$, and $333\,$nm are clearly visible. 
The CE position $x_{\rm CE}$ 
is the location of the central line 
and is measured with a typical accuracy of $3\,$$\mu$m.
\vspace{-5mm}}
\label{countrate}
\end{figure}

A typical Si $\mu$-strip count spectrum 
near the CE is shown in Fig.~\ref{countrate} 
for the multiline UV mode of the laser used in this measurement. 
This setting corresponds to 3 groups of lines 
centered around $364$, $351$, $333\,$nm, 
which are clearly resolved. 
The fitting function, 
also plotted, 
is based on the sum of 3 error functions 
plus background 
and includes 6 free parameters. 
The CE position, 
$x_{\rm CE}$, 
is taken as the location of the central line. 
The steep slope of the CE 
permits an excellent measurement of $x_{\rm CE}$
with a resolution of $\sim 3\,\mu$m 
for a statistics of about 10$^6$ counts. 
The specially designed VIRTEX-II electronics~\cite{VIRTEX} 
allowed us to collect such a statistics every $30\,$s, 
as compared to a few hours in our previous measurement.

\begin{figure}
\begin{center}
\includegraphics[width=0.9\hsize]{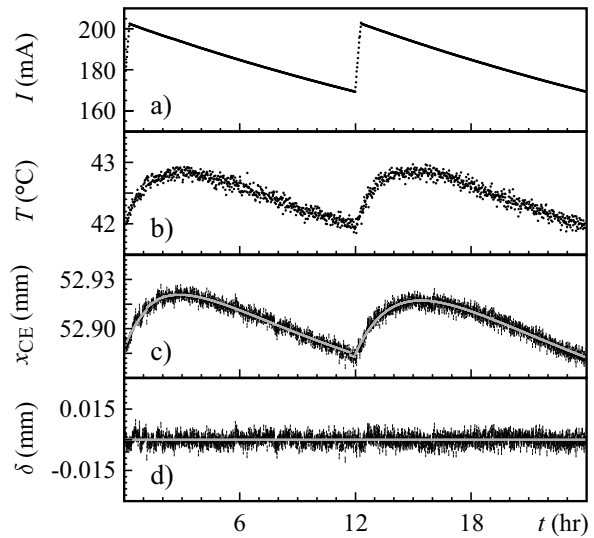}
\end{center}
\vskip-10pt
\caption{Time evolution over a day of 
a) ESRF beam intensity; 
b) tagging-box temperature; 
c) CE position and fitted curve (Eq.~\ref{FCE}); 
d) $\delta = x_{\rm CE} - x_{\rm fit}$. 
The error bars on position measurements are directly given by the CE fit.
\vspace{-5mm}}
\label{t_evol}
\end{figure}

During a week of data taking in July 2008, 
a total of 14765~CE 
spectra have been recorded. 
A sample of the time series of the CE positions 
relative to the ESRF beam 
covering $24\,$h 
is displayed in Fig.~\ref{t_evol}c, 
along with the tagging-box temperature (Fig.~\ref{t_evol}b) 
and the ESRF beam intensity (Fig.~\ref{t_evol}a). 
The sharp steps present in Fig.~\ref{t_evol}a 
correspond to the twice-a-day refills of the ESRF ring. 
The similarity of the temperature and CE spectra
combined with their correlation with the ESRF beam intensity 
led us to interpret the continuous and slow drift of the CE positions 
as a result of the tagging-box dilation.
To remove this trivial time dependence, 
we have fitted our raw data with the function
\beq{FCE}
x_{\rm{fit}}(t) = x_0 + a\left[1-\exp(- t/\tau_1)\right]\exp(-t/\tau_2)\,,
\eeq
where $t=0$ is chosen appropriately, 
and $\tau_1$, $\tau_2$ are time constants 
describing heat diffusion and ESRF beam-intensity decay, 
respectively. 
These two constants could be extracted directly 
from the temperature data (Fig.~\ref{t_evol}b).
Consequently, 
each time series between two refills ($12\,$h) 
has been fitted with only two free parameters: 
the position offset $x_0$ and the amplitude $a$.
The corrected and final spectrum, 
obtained by subtraction of the fitted function from the raw data, 
is plotted in Fig.~\ref{t_evol}d. 

This correction procedure is critical 
since its amplitude is 3 orders of magnitude larger 
than the extracted limit.
To test its validity, 
the same analysis has been applied to simulated data 
obtained by adding a known harmonic oscillation to the experimental data.
This study has shown that 
the signal amplitude would indeed be attenuated
by a factor ranging from 1.3 to 2.3 
depending on the phase.

The usual equation for the deflection of charges in a magnetic field
together with momentum conservation in Compton scattering
determines the relation between the CE position $x_{\rm CE}$ 
and the photon 3-momentum $\lambda'$:
\beq{XCE}
x_{\rm CE} = \frac{\lambda'}{p - \lambda'}\,C\,,
\eeq
where $p$ is again the momentum of the ESRF beam. 
The constant $C$ is determined 
by the trajectory of the electron; 
it therefore depends on the $B$-field magnitude and geometrical factors. 
Equation~\rf{XCE} can be used to derive the CE displacement resulting from
a change in $\lambda'$:
\beq{DLL}
\frac{\Delta x_{\rm CE}}{x_{\rm CE}} = 
\frac{p}{p - \lambda_{\rm CE}}\frac{\Delta \lambda'}{\lambda'}\,.
\eeq

\begin{figure}
\begin{center}
\includegraphics[width=0.9\linewidth]{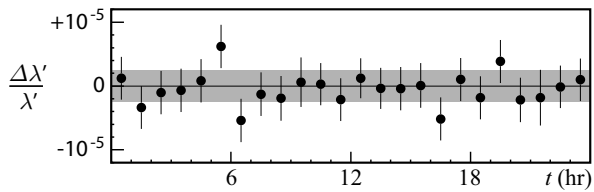}
\end{center}
\vskip-10pt
\caption{Full set of data folded modulo a sidereal day (24~bins). 
The error bars are purely statistical 
and agree with the dispersion of the data points ($\chi^2=1.04$ for the unbinned histogram).
The shaded area corresponds to the region of non-excluded signal amplitudes. \vspace{-5mm}}
\label{final}
\end{figure}

To search for a modulation, 
the 14765 data points have been folded modulo a sidereal day 
and divided in 24~bins (Fig.~\ref{final}). The error bars are purely statistical
and in agreement with a null signal ($\chi^2=1.04$ for the unbinned histogram). Moreover,
the histogram has been Fourier analyzed 
and provides no evidence for a harmonic oscillation 
at the sidereal frequency. 
Hence, 
an upper bound on a hypothetical oscillation can be extracted.
To exclude the signal hypothesis $A\sin(\Omega t + \phi)$ 
at a given confidence level (CL),
a statistical analysis has been developed.
In particular, 
we have chosen the
Bayesian approach \cite{PDG}, 
which allows us to take into account
the $\phi$-dependent attenuation factor
due to our correction procedure in a natural way. 
The p.d.f.\ for the amplitude
of the signal is generated by incorporating directly the attenuation factor
in the likelihood function before integration over $\phi$. The resulting
upper bound is $A<2.5\times10^{-6}$ 
at $95\%$ CL.

We next consider effects that 
could conceal an actual sidereal signal.  
Besides a direct oscillation of the orbit,
the two quantities that 
may affect the result appear in Eq.~\rf{XCE}: 
the dipole magnetic field  
via $C$, 
and the momentum of the ESRF beam $p$. 
All these parameters are linked to the machine operation, 
and their stability follows directly from the accelerator performance. 
At the ESRF, 
the electron orbit is precisely monitored to better than a $\mu$m, 
and the beam is maintained on its so-called golden orbit 
by adjustment of the RF-generator frequency.
Access to the accelerator database has allowed us 
to analyze the time series of both the dipole magnetic field 
and the orbit position close to our tagging system. 
We have verified that 
the fluctuations of either parameter 
do not exceed $10^{-6}$. 
The momentum stability of the ESRF beam,
given by $\int B\, dl$, 
is then ensured to be better than $10^{-6}$.
We therefore estimate that 
a sidereal oscillation related to one of the machine parameters 
cannot exceed a few parts in $10^{7}$ 
and is negligible.

We can now conclude that
our upper bound on a hypothetical sidereal oscillation of the CE energy is:
\beq{A_bound}
\Delta \lambda'/\lambda' < 2.5\times10^{-6}\quad (95\,\% \; {\rm CL})\,,
\eeq
yielding the competitive limit
$\sqrt{\kappa_X^2+\kappa_Y^2} < 1.6\times10^{-14}$ ($95\,\%$ CL)
with Eq.~\rf{sidereal}.
Reinstating mSME notation and the electron coefficients
for generality gives at $95\,\%$ CL
\beq{Bound}
{}\hspace{-.3mm}\sqrt{[2c_{TX}-(\kop)^{YZ}]^2+[2c_{TY}-(\kop)^{ZX}]^2} 
< 1.6\times10^{-14}
\eeq
improving previous bounds by a factor of ten. 
The other, omitted mSME coefficients leave
this limit unaffected.

In summary, 
we have made use of the GRAAL $\gamma$-ray beam 
produced by inverse Compton scattering 
off the high-energy electrons circulating in the ESRF ring 
to test the isotropy of light propagation. 
This represents the first test of Special Relativity 
via a non-threshold kinematics effect in a particle collision.  
Our measurement, 
based on the search for sidereal modulations 
in the CE of the scattered electrons, 
shows no evidence for a signal. 
Interpreting our result within the mSME, 
we have obtained a competitive upper bound on a combination of coefficients 
of the mSME's QED sector 
at the $10^{-14}$ level.

\acknowledgments 

It is a pleasure to thank the ESRF staff 
for the smooth operation of the storage ring. 
We are especially indebted to L.~Farvacque and J.M.~Chaize 
for fruitful discussions and access to the machine database. 
Contributions from G.~Angeloni (VIRTEX-II electronics), 
G.~Nobili (technical support), 
and O. Zimmermann (laser stabilization loop) 
are greatly appreciated.
R.~Lehnert acknowledges support from CONACyT under Grant No.\ 55310.

\end{document}